\documentclass[final,5p,times,twocolumn]{elsarticle}
\usepackage{lineno,hyperref,amsmath,amsthm,amssymb,amsfonts,ragged2e,color,subfig}
\journal{``Pramana-Journal of Physics"}
\begin{document}
\begin{frontmatter}
\title{Dust-acoustic rogue waves in non-thermal plasmas}
\author{S. K. Paul$^{*1}$, N. A. Chowdhury$^{**,2,3}$, A. Mannan$^{\dag,2,4}$, and A. A. Mamun$^{\ddag,2}$}
\address{$^1$Department of Basic Sciences and Humanities, University of Asia Pacific, Farmgate, Dhaka-1205, Bangladesh\\
$^2$Department of Physics, Jahangirnagar University, Savar, Dhaka-1342, Bangladesh\\
$^3$Plasma Physics Division, Atomic Energy Centre, Dhaka-1000, Bangladesh\\
$^4$Institut f{\"u}r Mathematik, Martin Luther Universit{\"a}t Halle-Wittenberg, D-06099 Halle, Germany\\
e-mail: $^*$sanjitkpaul@gmail.com, $^{**}$nurealam1743phy@gmail.com,\\
$^{\dag}$abdulmannan@juniv.edu, $^\ddag$mamun\_phys@juniv.edu,}
\begin{abstract}
 The nonlinear propagation of dust-acoustic (DA) waves (DAWs) and associated DA rogue waves (DARWs),
  which are governed by the nonlinear Schr\"{o}dinger equation, is theoretically investigated
  in a four component plasma medium containing inertial warm negatively charged dust grains and
  inertialess non-thermal distributed electrons as well as iso-thermal positrons and ions.
  The modulationally  stable and unstable parametric regimes of DAWs are numerically studied for the plasma parameters.
  Furthermore, the effects of temperature ratios of ion-to-electron and ion-to-positron, and the number density
  of ion and dust grains on the DARWs are investigated. It is observed that the physical parameters play a very crucial role in
  the formation of DARWs. These results may be useful in understanding the electrostatic
  excitations in dusty plasmas in space and laboratory situations.
\end{abstract}
\end{frontmatter}
\section{Introduction}
\label{1sec:Introduction}
The research regarding the propagation of nonlinear electrostatic perturbation in a four
component electron-positron-ion-dust (FCEPID) plasma medium (FCEPIDPM) has received a great attention
among the plasma physicists due to the existence of FCEPID not only in space plasmas,
viz., the hot spots on dust rings in the galactic centre \cite{Paul2017,Banerjee2016,Jehan2009,Saberian2017},
auroral zone \cite{Banerjee2016}, around  pulsars \cite{Jehan2009,Esfandyari-Kalejahi2012},
interstellar medium \cite{Sardar2016}, Milky Way \cite{Sardar2016}, accretion disks near neutron stars \cite{Sardar2016},
Jupiter's magnetosphere \cite{Paul2016} but also in laboratory experiments. A number of
authors have studied dust-acoustic (DA) waves (DAWs) \cite{Jehan2009,Saberian2017,Esfandyari-Kalejahi2012},
dust-ion-acoustic waves (DIAWs) \cite{Paul2017,Banerjee2016}, and associated nonlinear structures
in four component dusty plasma. The presence of positron drastically changes the mechanism of the
formation and propagation of nonlinear electrostatic structures, viz., shock, solitons \cite{Banerjee2016},
super-solitons \cite{Paul2017,Paul2016}, and  double layers (DLs) \cite{Banerjee2016}, etc.

The existence of the non-thermal particles (viz., electrons \cite{Paul2017,Banerjee2016,Sardar2016,Paul2016},
positrons \cite{Paul2017,Banerjee2016}, and ions, etc.), which rigorously change the dynamics of the plasma
medium and the mechanism of the formation of various electrostatic pulses,
in space plasmas has confirmed by the Viking \cite{Bostrom1992} and Freja satellites \cite{Dovner1994}.
Cairns \textit{et al.} \cite{Cairns1995} investigated the electrostatic solitary structures in a
non-thermal plasma, and found that both negative and positive density perturbations can exist
in the presence of non-thermal electrons.  Paul \textit{et al.} \cite{Paul2017}
examined that the existence of non-thermal electrons and positrons in a FCEPID model supports the solitary waves of both
polarities. Banerjee and Maitra \cite{Banerjee2016} studied DIAWs, i.e., solitons and DLs, in a FCEPIDPM
having non-thermal electrons and positrons and observed that the  positive
potential can not exist after a particular value of $\alpha$.

The rogue waves (RWs), which appear due to the modulational instability (MI) of the carrier waves,
have been observed in different branch of sciences, viz., Oceanography \cite{Kharif2009},
Biology, Optics \cite{Solli2007}, Finance \cite{Yan2010},
Plasma physics \cite{Chowdhury2018a,Chowdhury2018b,Kourakis2003,Rahman2018},
and  are also governed by the nonlinear Schr\"{o}dinger
equation (NLSE) \cite{Chowdhury2018a,Chowdhury2018b,Kourakis2003,Rahman2018}.
Chowdhury \textit{et al.} \cite{Chowdhury2018b} examined the stability conditions for the positron-acoustic waves (PAWs)
in a multi-component plasma medium, and found that the critical wave number ($k_c$), which indicates the
stable and unstable parametric regimes of PAWs, reduces with the increase in the value
of $\alpha$. Kourakis and Shukla \cite{Kourakis2003}
studied the MI of the DIAWs in a three component plasma medium and showed that
the existence of the negative dust grains reduces the value of $k_c$.
Rahman \textit{et al.} \cite{Rahman2018} reported the stable and unstable domains in a dusty plasma
medium having non-thermal plasma species, and found that the amplitude of the DA RWs (DARWs)
increases with non-thermality of the plasma species.

Recently, Saberian \textit{et al.} \cite{Saberian2017} studied the DA solitons and DLs in a FCEPIDPM.
Esfandyari-Kalejahi \textit{et al.} \cite{Esfandyari-Kalejahi2012} investigated DA solitary (DASWs)
in a multi-component dusty plasma in presence of non-thermal electrons, and found that the amplitude of the
DASWs increases with charge state of the negative dust grains.
Jehan \textit{et al.} \cite{Jehan2009} considered an unmagnetized FCEPIDPM having
negatively charged massive inertial dust grains, inertialess electrons, positrons, and ions for
examining DASWs, and observed that the existence of positrons and it's temperature
can change the sign of the nonlinear coefficient of the  governing equation. The aim of the present paper is
to further significant amount of extension of Jehan \textit{et al.} \cite{Jehan2009} work by studying the MI
of the DAWs in an unmagnetized FCEPIDPM having non-thermal electrons
featuring Cairns' distribution, and also examine the nonlinear properties of DARWs.

The manuscript is organized as follows: The basic governing equations of our
plasma model are presented in Sec. \ref{1sec:Governing Equations}.  The MI and RWs
are presented in Sec. \ref{1sec:Modulational instability and  Rogue waves}. Results and
discussion are provided in Sec. \ref{1sec:Results and discussion}.
The conclusion is  provided in Sec. \ref{1sec:Conclusion}.
\section{Governing Equations}
\label{1sec:Governing Equations}
We consider the propagation of DAWs in an unmagnetized collisionless FCEPIDPM
consisting of inertial warm negatively charged massive dust grains (mass $m_d$;
charge $q_d=-Z_de$), and inertialess non-thermal Cairns' distributed electrons
(mass $m_e$; charge $q_e=-e$) as well as iso-thermal  positrons (mass $m_p$; charge $q_p=+e$)
and ions (mass $m_i$; charge $q_i=+Z_ie$), where $Z_d$ ($Z_i$) is the number of
electron (proton) residing on a negatively (positively) charged massive dust grains (ions).
Overall, the charge neutrality condition for our plasma model can be written as
$n_{e0} +Z_{d} n_{d0}= n_{p0}+Z_in_{i0}$. Now, the basic set of normalized equations can be written as
\begin{eqnarray}
&&\hspace*{-1.3cm}\frac{\partial n_d}{\partial t}+\frac{\partial(n_d u_d)}{\partial x}=0,
\label{1eq:1}\\
&&\hspace*{-1.3cm}\frac{\partial u_d}{\partial t} + u_d\frac{\partial u_d}{\partial x}+\sigma_1 n_d\frac{\partial n_d}{\partial x}=\frac{\partial \phi}{\partial x},
\label{1eq:2}\\
&&\hspace*{-1.3cm}\frac{\partial^2 \phi}{\partial x^2}=\mu_e n_e-(1+\mu_e-\mu_i)n_p-\mu_i n_i + n_d,
\label{1eq:3}\
\end{eqnarray}
where $n_d$ is the number density of warm dust grains normalized by its equilibrium value $n_{d0}$;
$u_d$ is the dust fluid speed normalized by the DA wave speed $C_d=(Z_dk_BT_i/m_d)^{1/2}$
(with $T_i$ being the ion temperature, $m_d$ being the dust grain mass, and $k_B$ being the Boltzmann
constant); $\phi$ is the electrostatic wave potential normalized by $k_BT_i/e$ (with $e$ being the
magnitude of single electron charge); the time and space variables are normalized by
$\omega_{pd}^{-1}=(m_d/4\pi Z_d^2e^2n_{d0})^{1/2}$ and $\lambda_{Dd}=(k_BT_i/4\pi Z_dn_{d0}e^2)^{1/2}$,
respectively; $P_d=P_{d0}(N_d/n_{d0})^\gamma$ [with $P_{d0}$ being the equilibrium adiabatic
pressure of the dust, and $\gamma=(N+2)/N$, where $N$ is the degree of freedom and for one-dimensional case,
$N=1$ then $\gamma=3$]; $P_{d0}=n_{d0}k_BT_d$ (with $T_d$ being the temperatures of the warm dust grains);
and other plasma parameters are considered as $\sigma_1=3T_d/Z_dT_i$, $\mu_e=n_{e0}/Z_{d}n_{d0}$, and $\mu_i=Z_{i}n_{i0}/Z_{d}n_{d0}$.
Now, the expression for the number density of  non-thermal electrons following the
Cairns' distribution \cite{Paul2017,Cairns1995} can be written as
\begin{eqnarray}
&&\hspace*{-1.3cm}n_e= [1-\beta\sigma_{2}\phi +\beta\sigma_{2}^{2}\phi^{2}]\exp(\sigma_2\phi),
\label{1eq:4}\
\end{eqnarray}
where  $\sigma_2=T_i/T_e$ ($T_e$ being the temperature of the iso-thermal electron), and
$\beta=4\alpha/(1+3\alpha)$ with $\alpha$ being the parameter determining the fast
particles present in our plasma model.
Now, the expression for the number density of  iso-thermal positrons following the
Maxwellian distribution can be written as \cite{Jehan2009}
\begin{eqnarray}
&&\hspace*{-1.3cm}n_p= \exp(-\sigma_{3} \phi),
\label{1eq:5}\
\end{eqnarray}
where $\sigma_3=T_i/T_p$ ($T_p$ being the temperature of the iso-thermal positrons).
Now, the expression for the number density of  iso-thermal ions following the
Maxwellian distribution  can be written as \cite{Jehan2009}
\begin{eqnarray}
&&\hspace*{-1.3cm}n_i=\exp(-\phi).
\label{1eq:6}\
\end{eqnarray}
Now, by substituting Eqs. \eqref{1eq:4}-\eqref{1eq:6} in  Eq. \eqref{1eq:3}, and expanding up to third order of $\phi$, we get
\begin{eqnarray}
&&\hspace*{-1.3cm}\frac{\partial^2 \phi}{\partial x^2}+1=n_d+S_1\phi+ S_2\phi^2+ S_3\phi^3+...
\label{1eq:7}\
\end{eqnarray}
where
\begin{eqnarray}
&&\hspace*{-1.3cm}S_1=(\sigma_2-\beta\sigma_2+\sigma_3)\mu_e +(1-\sigma_3)\mu_i+\sigma_3,
\nonumber\\
&&\hspace*{-1.3cm}S_2=[{(\sigma_2^{2}-\sigma_3^{2})\mu_e +(\sigma_3^{2}-1)\mu_i-\sigma_3^{2}}]/{2},
\nonumber\\
&&\hspace*{-1.3cm}S_3=[{(3\beta+1)\mu_e \sigma_2^{3} +\mu_e \sigma_3^{3}+(1-\sigma_3^{3})\mu_i +\sigma_3^{3}}]/{6}.
\nonumber\
\end{eqnarray}
The terms containing $S_1$, $S_2$, and $S_3$ in the right hand side of
Eq. \eqref{1eq:7} are the contribution of inertialess electrons, positrons, and ions. We note
that Eqs. \eqref{1eq:1},  \eqref{1eq:2}, and  \eqref{1eq:7} now represent the basis set of
normalized equations to describe the nonlinear dynamics of the DAWs, and associated DARWs in
an unmagnetized FCEPIDPM under consideration.
\section{Derivation of the NLSE}
\label{1sec:Derivation of the NLSE}
To study the MI of the DAWs, we want to derive the NLSE by employing the reductive perturbation method,
and for that case, first we can write the stretched co-ordinates in the
form \cite{Chowdhury2018a,Chowdhury2018b,Kourakis2003,Rahman2018}
\begin{eqnarray}
&&\hspace*{-1.3cm}\xi={\epsilon}(x-v_g t),
\label{1eq:8}\\
&&\hspace*{-1.3cm}\tau={\epsilon}^2 t,
\label{1eq:9}\
\end{eqnarray}
where $v_g$ is the group speed and $\epsilon$ is a small parameter. Then we can write the dependent variables as
\begin{eqnarray}
&&\hspace*{-1.3cm}n_{d}=1 +\sum_{m=1}^{\infty}\epsilon^{m}\sum_{l=-\infty}^{\infty}n_{dl}^{(m)}(\xi,\tau)~\mbox{exp}[i l(kx-\omega t)],
\label{1eq:10}\\
&&\hspace*{-1.3cm}u_{d}=\sum_{m=1}^{\infty}\epsilon^{m}\sum_{l=-\infty}^{\infty}u_{dl}^{(m)}(\xi,\tau)~\mbox{exp}[i l(kx-\omega t)],
\label{1eq:11}\\
&&\hspace*{-1.3cm}\phi=\sum_{m=1}^{\infty}\epsilon^{m}\sum_{l=-\infty}^{\infty}\phi_{l}^{(m)}(\xi,\tau)~\mbox{exp}[i l(kx-\omega t)],
\label{1eq:12}
\end{eqnarray}
where $k$ and $\omega$ are real variables representing the carrier wave number and frequency, respectively.
The derivative operators can be written as \cite{Chowdhury2018a,Chowdhury2018b,Kourakis2003,Rahman2018}
\begin{eqnarray}
&&\hspace*{-1.3cm}\frac{\partial}{\partial t}\rightarrow\frac{\partial}{\partial t}-\epsilon v_g\frac{\partial}{\partial\xi}+\epsilon^2\frac{\partial}{\partial\tau},
\label{1eq:13}\\
&&\hspace*{-1.3cm}\frac{\partial}{\partial x}\rightarrow\frac{\partial}{\partial x}+\epsilon\frac{\partial}{\partial\xi}.
\label{1eq:14}
\end{eqnarray}
Now, by substituting Eqs. \eqref{1eq:8}-\eqref{1eq:14}  into  Eqs. \eqref{1eq:1}- \eqref{1eq:2}, and \eqref{1eq:7}, and
collecting the terms containing $\epsilon$, the first order ($m=1$ with $l=1$)  reduced equations can be written as
\begin{eqnarray}
&&\hspace*{-1.3cm}n_{d1}^{(1)}=\frac{k^2}{S}\phi_1^{(1)},
\label{1eq:15}\\
&&\hspace*{-1.3cm}u_{d1}^{(1)}=\frac{k \omega}{S}\phi_1^{(1)},
\label{1eq:16}\\
&&\hspace*{-1.3cm}n_{d1}^{(1)}= -k^2\phi_1^{(1)}-S_1 \phi_1^{(1)},
\label{1eq:17}\
\end{eqnarray}
where $S=\sigma_1 k^2-\omega^2 $. Hence these relations provide the dispersion relation of DAWs
\begin{eqnarray}
&&\hspace*{-1.3cm}\omega^2=\sigma_1 k^2+\frac{k^2}{S_1+k^2}.
\label{1eq:18}\
\end{eqnarray}
The second-order ($m=2$ with $l=1$) equations are given by
\begin{eqnarray}
&&\hspace*{-1.3cm}n_{d1}^{(2)}=\frac{k^2}{S}\phi_1^{(2)}-\frac{2ik\omega(v_g k-\omega)}{S^2} \frac{\partial \phi_1^{(1)}}{\partial\xi},
\label{1eq:19}\\
&&\hspace*{-1.3cm}u_{d1}^{(2)}=\frac{k \omega}{ S}\phi_1^{(2)} -\frac{i(v_g k-\omega)(\omega^2 + \sigma_1 k^2 )}{S^2} \frac{\partial \phi_1^{(1)}}{\partial\xi},
\label{1eq:20}\
\end{eqnarray}
with the compatibility condition
\begin{eqnarray}
&&\hspace*{-1.3cm}v_g=\frac{\partial \omega}{\partial k}=\frac{\omega^2-S^2}{k\omega}.
\label{1eq:21}\
\end{eqnarray}
The coefficients of $\epsilon$ for $m=2$ and $l=2$ provide the second order harmonic
amplitudes which are found to be proportional to $|\phi_1^{(1)}|^2$
\begin{eqnarray}
&&\hspace*{-1.3cm}n_{d2}^{(2)}=S_4|\phi_1^{(1)}|^2,
\label{1eq:22}\\
&&\hspace*{-1.3cm}u_{d2}^{(2)}=S_5 |\phi_1^{(1)}|^2,
\label{1eq:23}\\
&&\hspace*{-1.3cm}\phi_{2}^{(2)}=S_6 |\phi_1^{(1)}|^2,
\label{1eq:24}\
\end{eqnarray}
where
\begin{eqnarray}
&&\hspace*{-1.3cm}S_4=\frac{2S_6 k^2 S^2 -\sigma_1 k^6  -3 \omega^2 k^4  }{2S^3},
\nonumber\\
&&\hspace*{-1.3cm}S_5=\frac{ S_4 \omega  S^2  -\omega k^4 }{k S^2},
\nonumber\\
&&\hspace*{-1.3cm}S_6=\frac{\sigma_1 k^6 + 3\omega^2 k^4 -2 S_2 S^3}{6k^2 S^3}.
\nonumber\
\end{eqnarray}
Now, we consider the expression for ($m=3$ with $l=0$) and ($m=2$ with $l=0$),
which leads the zeroth harmonic modes. Thus, we obtain
\begin{eqnarray}
&&\hspace*{-1.3cm}n_{d0}^{(2)}=S_{7}|\phi_1^{(1)}|^2,
\label{1eq:25}\\
&&\hspace*{-1.3cm}u_{d0}^{(2)}=S_{8}|\phi_1^{(1)}|^2,
\label{1eq:26}\\
&&\hspace*{-1.3cm}\phi_0^{(2)}=S_{9} |\phi_1^{(1)}|^2,
\label{1eq:27}\
\end{eqnarray}
where
\begin{eqnarray}
&&\hspace*{-1.3cm}S_{7}=\frac{S_9 S^2-\sigma_1 k^4-\omega^2 k^2-2\omega v_g k^3}{S^2(\sigma_1-v^2_g)},
\nonumber\\
&&\hspace*{-1.3cm}S_{8}=\frac{S_{7} v_g  S^2-2\omega k^3 }{S^2},
\nonumber\\
&&\hspace*{-1.3cm}S_{9}=\frac{2\omega v_g k^3 +\sigma_1  k^4+\omega^2 k^2-2S_{2} S^2(\sigma_1-v^2_g)}{ S^2(1-S_1 v^2_g +\sigma_1S_1)}.
\nonumber\
\end{eqnarray}
Finally, the third harmonic modes ($m=3$) and ($l=1$), with the help of \eqref{1eq:15}-\eqref{1eq:27},
give a set of equations, which can be reduced to the following  NLSE:
\begin{eqnarray}
&&\hspace*{-1.3cm}i\frac{\partial\Phi}{\partial\tau}+P\frac{\partial^2\Phi}{\partial\xi^2}+Q|\Phi|^2\Phi=0,
\label{1eq:28}
\end{eqnarray}
where $\Phi=\phi_1^{(1)}$ for simplicity. In Eq. \eqref{1eq:28}, the dispersion coefficient $P$ is as
\begin{eqnarray}
&&\hspace*{-1.3cm}P=\frac{F_{1}(v_g k -\omega)-S^3 }{2\omega k^2 S}.
\nonumber\
\end{eqnarray}
where $ F_{1}=\omega^3+3\omega\sigma_1k^2-3v_gk\omega^2-v_g\sigma_1k^3$,
and also in Eq. \eqref{1eq:28}, the nonlinear coefficient $Q$ is as
\begin{eqnarray}
&&\hspace*{-1.3cm}Q=\frac{2S_2S^2(S_9+S_6)+3S_3 S^2-F_2}{2\omega k^2}.
\nonumber\
\end{eqnarray}
\begin{figure}
\centering
\includegraphics[width=80mm]{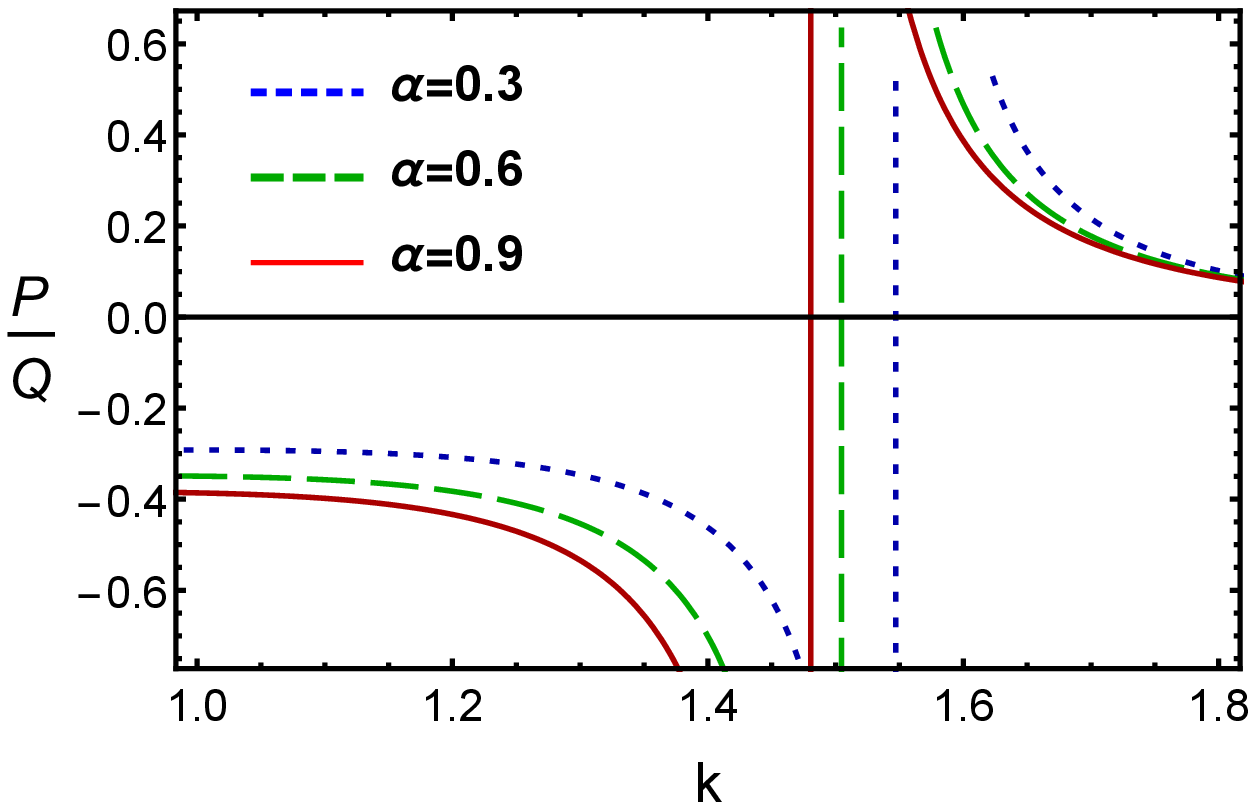}
\caption{Plot of $P/Q$ vs $k$ for different values of $\alpha$ along with $\mu_e=0.5$, $\mu_i=0.7$, $\sigma_1=0.003$, $\sigma_2=0.4$, and $\sigma_3=0.5$.}
\label{1Fig:F1}
\end{figure}
\begin{figure}
\centering
\includegraphics[width=80mm]{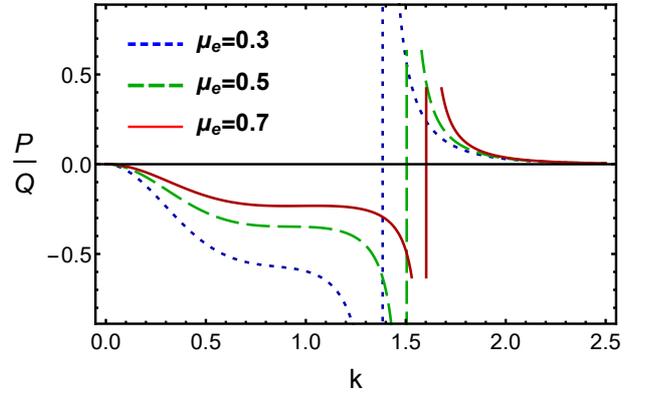}
\caption{Plot of $P/Q$ vs $k$ for different values of $\mu_e$ along with  $\alpha=0.6$, $\mu_i=0.7$, $\sigma_1=0.003$, $\sigma_2=0.4$, and $\sigma_3=0.5$.}
\label{1Fig:F2}
\end{figure}
where $F_2= (k^4\sigma_1+\omega^2k^2)(S_4+S_7)+ 2\omega k^3(S_5+S_8)$.
The space and time evolution of the DAWs in an unmagnetized FCEPIDPM are directly governed by the
coefficients $P$ and $Q$, and indirectly governed by different plasma parameters
such as $\alpha$, $\mu_i$,  $\mu_e$, $\sigma_1$, $\sigma_2$, $\sigma_3$, and $k$. Thus,
these plasma parameters significantly affect the stability conditions of the DAWs in an unmagnetized FCEPIDPM.
\section{Modulational instability and  Rogue waves}
\label{1sec:Modulational instability and  Rogue waves}
The stable and unstable parametric regimes of DAWs are organized by the sign of $P$ and $Q$ of
Eq. \eqref{1eq:28} \cite{Chowdhury2018a,Chowdhury2018b,Kourakis2003,Rahman2018,Fedele2002,Sun2018,Chai2019}.
When $P$ and $Q$ have same sign (i.e., $P/Q>0$), the evolution of the DAWs amplitude is
modulationally unstable in presence of the external perturbations. On the other hand,
when $P$ and $Q$ have opposite sign (i.e., $P/Q<0$), the DAWs are modulationally stable in presence of the external perturbations.
The plot of $P/Q$ against $k$ yields stable and unstable parametric regimes of the DAWs.
The point, at which the transition of $P/Q$ curve intersect with $k$-axis, is known as threshold
or critical wave number $k$ ($=k_c$) \cite{Fedele2002,Sun2018,Chai2019,C1,C2,C3,C4,C5}.
\begin{figure}
\centering
\includegraphics[width=80mm]{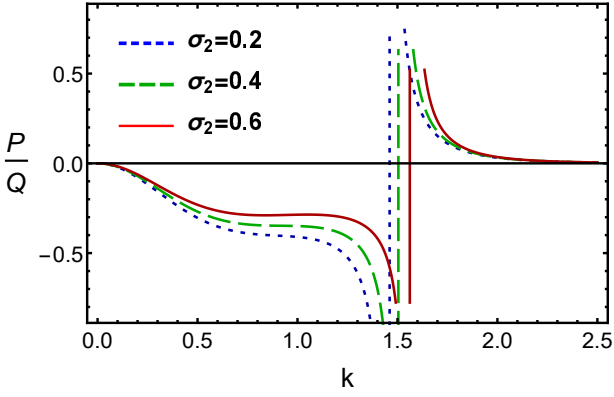}
\caption{Plot of $P/Q$ vs $k$ for different values of $\sigma_2$, along with $\alpha=0.6$,  $\mu_e=0.5$, $\mu_i=0.7$, $\sigma_1=0.003$,  and $\sigma_3=0.5$.}
\label{1Fig:F3}
\end{figure}
\\
The governing equation for highly energetic DARWs in the modulationally unstable
parametric regime ($P/Q>0$) can be written as \cite{Akhmediev2009,Ankiewiez2009,C6,C7,C8}
\begin{eqnarray}
&&\hspace*{-1.3cm}\Phi(\xi,\tau)= \sqrt{\frac{2P}{Q}} \left[\frac{4(1+4iP\tau)}{1+16P^2\tau^2+4\xi^2}-1 \right]\mbox{exp}(2iP\tau).
\label{1eq:29}
\end{eqnarray}
\begin{figure}
\centering
\includegraphics[width=80mm]{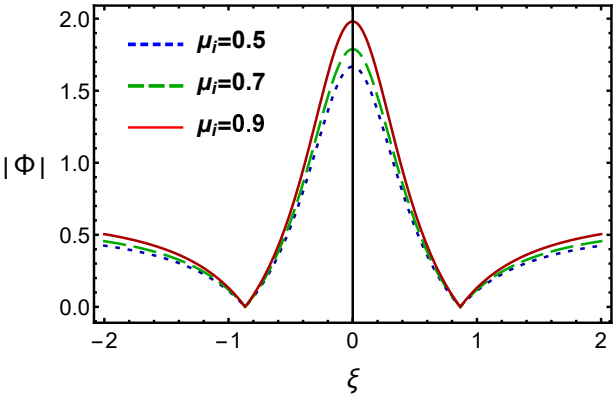}
\caption{Plot of $|\Phi|$ vs $\xi$ for different values of $\mu_i$, along with $k=1.7$, $\tau=0$, $\alpha=0.6$  $\mu_e=0.5$, $\sigma_1=0.003$, $\sigma_2=0.4$, and $\sigma_3=0.5$.}
\label{1Fig:F4}
\end{figure}
Equation \eqref{1eq:29} describes that a large amount of wave energy, which causes due to the
nonlinear characteristics of the plasma medium, is localized into a comparatively small area in space.
\section{Results and discussion}
\label{1sec:Results and discussion}
We have numerically analyzed the stable and unstable parametric regimes of DAWs in
Figs. \ref{1Fig:F1}-\ref{1Fig:F3}. The effects of non-thermality
of the electrons in organizing the stable and unstable parametric regimes
of the DAWs can be seen from Fig. \ref{1Fig:F1} which indicates that
the $k_c$ as well as modulationally stable parametric regime of DAWs decreases
with the increase in the value of $\alpha$, and this result is a good agreement with the
result of Ref. \cite{Chowdhury2018b}. So, the DAWs become modulationally unstable for small values of $k$
for excess non-thermality of the plasma species.

Figure \ref{1Fig:F2} indicates that the presence of the non-thermal
electrons and negatively charged warm dust grains can significantly
modify the stability conditions of DAWs in an unmagnetized FCEPIDPM.
It is clear from this figure that (a) when $\mu_e=0.3$, $0.5$, and
$0.7$ then the corresponding $k_c$ value is $k_c\equiv1.4$ (dotted blue curve),
$k_c\equiv1.5$ (dashed green curve), and $k_c\equiv1.6$ (solid red curve);
(b) the critical value increases with the increase in the value of the $\mu_e$;
(c) the modulationally stable (unstable) parametric regime of DAWs
increases with an increase in the value of electron (dust) number
density for constant value of the charge state of negative dust
grains (via $\mu_e$).
\begin{figure}
\centering
\includegraphics[width=80mm]{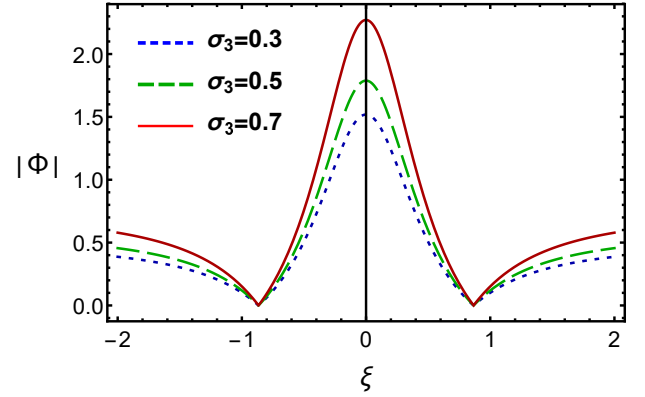}
\caption{Plot of $|\Phi|$ vs $\xi$ for different values of $\sigma_3$ along with  $k=1.7$, $\tau=0$,  $\alpha=0.6$  $\mu_e=0.5$, $\mu_i=0.7$, $\sigma_1=0.003$, and $\sigma_2=0.4$.}
\label{1Fig:F5}
\end{figure}
\\
The effects of ion to electron temperature (via $\sigma_2$) on the stable and unstable
parametric regimes can be observed in Fig. \ref{1Fig:F3}, and it is clear from this
figure that (a) the $k_c$ increases with  $\sigma_2$; (b) the modulationally stable
parametric regime increases with ion temperature while decreases with electron temperature.
So, the temperature of the electron and ion plays an opposite role in recognizing the
modulationally stable and unstable parametric regimes of DAWs in an unmagnetized FCEPIDPM.

We have numerically analyzed Eq. \eqref{1eq:29} in Figs. \ref{1Fig:F4}  and
\ref{1Fig:F5} to understand the nonlinear property of the FCEPIDPM
as well as the mechanism of the formation of DARWs associated with DAWs in the
modulationally unstable parametric regime. Figure \ref{1Fig:F4} indicates
that (a) the amplitude and width of the DARWs increase with an increase
in the value of $\mu_i$; (b) the nonlinearity, which organizes the shape
of the DARWs, of the plasma medium increases with an increase in the
value of ion number density whereas the nonlinearity of the plasma medium
decreases with  dust number density when their charge
state remain constant. Figure \ref{1Fig:F5} indicates the temperature
effects of ion and electron (via $\sigma_3$) on the generation of DARWs
in FCEPIDPM. The ion (positron) temperature enhances (suppresses) both
amplitude and width of the DARWs associated with DAWs in the modulationally
unstable parametric regime.
\section{Conclusion}
\label{1sec:Conclusion}
In this study, we have performed a nonlinear analysis of DAWs in an unmagnetized FCEPIDPM
consisting of inertial negatively charged warm dust grains,
and inertialess non-thermal Cairns' distributed electrons as well as iso-thermal positrons
and ions. The evolution of DAWs is governed by the standard NLSE, and
the coefficients $P$ and  $Q$ of NLSE can recognize the modulationally stable and
unstable parametric regimes of  DAWs in the presence of the external perturbation. It is observed that
the DAWs become unstable for small values of $k$ for excess non-thermality (via $\alpha$)
of the plasma species, and the stable (unstable) parametric regimes of the DAWs
increases with the increase in the value of electron (dust) number density
at constant value of the dust charge state. The ion (positron) temperature enhances (suppresses)
both amplitude and width of the DARWs. Finally, these results may be applicable
in understanding the conditions of the MI of DAWs and associated DARWs in astrophysical
environments (viz., the hot spots on dust rings in the galactic centre \cite{Paul2017,Banerjee2016,Jehan2009,Saberian2017},
auroral zone \cite{Banerjee2016}, around  pulsars \cite{Jehan2009,Esfandyari-Kalejahi2012},
interstellar medium \cite{Sardar2016}, Milky Way \cite{Sardar2016}, accretion disks near neutron stars \cite{Sardar2016},
and Jupiter's magnetosphere \cite{Paul2016}, etc) and laboratory devices.

\end{document}